\documentclass[prd,aps]{revtex4}

\usepackage{epsfig}
\newcommand{\beq}{\begin{equation}}
\newcommand{\eeq}{\end{equation}}
\newcommand{\be}{\begin{eqnarray}}
\newcommand{\ee}{\end{eqnarray}}

\newcommand{\bd}{B_d-\bar{B}_d}
\newcommand{\bs}{B_s-\bar{B}_s}
\newcommand{\bq}{B_q-\bar{B}_q}

\begin{document}

%\draft
\preprint{
\begin{tabular}{l}
\hbox to\hsize{June, 2008\hfill KIAS-P08060}\\[-2mm]
%\hbox to\hsize{              \hfill KAIST-TH 02/04}\\[-3mm]
%\hbox to\hsize{           \hfill hep-ph/yymmdd}\\[5mm] 
\end{tabular}
}

\title{
$B_d-\bar{B}_d$ mixing vs. $B_s-\bar{B}_s$ mixing
with the anomalous $Wtb$ couplings 
}

\author{ Jong Phil Lee }
%\email{jplee@kias.re.kr}

\affiliation{
School of Physics, Korea Institute for Advanced Study, Seoul 130-722, Korea
}

\author{ Kang Young Lee }
\email{kylee@muon.kaist.ac.kr}

\affiliation{
Department of Physics, Korea University, Seoul 136-713, Korea 
}

\date{\today}
%\maketitle

\begin{abstract}

We explore the effects of the anomalous $tbW$ couplings
on the $\bd$ mixing and recently measured $\bs$ mixing. 
The combined analysis of mixings via box diagrams with penguin decays
provides strong constraints on the anomalous top quark couplings.
We find the bound from the $\bd$ mixing data is stronger than
that from the $\bs$ mixing. 

\end{abstract}

%\pacs{PACS numbers:12.90.+b,13.20.He,14.65.Ha}
\maketitle

%\tightenlines

               %%%%%%%%%%%%%%%%%%%%%%%%%%%%%%%
               %%    BEGINNING OF TEXT      %%
               %%%%%%%%%%%%%%%%%%%%%%%%%%%%%%%

\section{Introduction}

Expected is the production of a large number of top quark pairs 
at the CERN Large Hadron Collider (LHC), 
which allows us to probe the top quark couplings 
\cite{lhc,toplhc}. 
The $tbW$ coupling will be directly tested with high precision 
through the dominant $t \to b W$ decays at the LHC
and other top decay channels are highly suppressed by small mixing angles. 
The present value of $|V_{tb}|$ is determined at the Tevatron to be
$|V_{tb}|>0.78$ \cite{vtbcdf} and $|V_{tb}|>0.89$ \cite{vtbd0}
at 95$\%$ C. L.
with assuming the the Cabibbo-Kobayashi-Maskawa (CKM) unitarity.
The direct determination of $|V_{tb}|$ without assuming unitarity
is performed through the single top quark production
and the CDF \cite{stcdf} and the D0 \cite{std0}
obtained the limit $|V_{tb}|>0.74$ at 95$\%$ C. L.

The standard model (SM) of electroweak and strong interactions 
has been successful in describing a wide range of experimental data so far.
The only unobserved ingredient of the SM is the Higgs boson
and a few coupling constants are not precisely tested yet.
The present measurements on $|V_{tb}|$ is still far from the SM
prediction $\sim 1$ from the CKM unitarity,
and there are still rooms for new physics (NP) beyond the SM
in the $tbW$ couplings.
Therefore it is very exciting to probe the new physics signature 
with the top quark couplings.
One of the most promising way to test the effects of the NP
in top quark couplings before the LHC is 
to study the neutral $B_q$ $(q=d,s)$ meson mixings.
The $\bq$ mixing arises through the box diagrams
with internal lines of $W$ boson and $u$-type quarks in the SM.
Since the top quark loop dominates,
the box diagrams are sensitive to the anomalous top couplings.  
The current average of the $\bd$ mixing is found to be
\cite{HFAG}
\be
\Delta M_d 
= ( 0.507 \pm 0.005 )~ {\rm ps}^{-1}.
\ee
Recently the measurements of the $\bs$ mixing
by the CDF \cite{cdf} and D0 \cite{d0} collaborations 
are reported to be 
\be
\Delta M_s 
%\equiv M_H^s - M_L^s = 2 | M_{12}^s | 
&=& ( 17.77 \pm 0.10 \pm 0.07 )~ {\rm ps}^{-1}
~~~~~~~~~~~~~~~~~~~({\rm CDF}), 
\nonumber \\ 
&=& ( 18.53 \pm 0.93 \pm 0.30 )~ {\rm ps}^{-1} 
~~~~~~~~~~~~~~~~~~~~ ({\rm D0}),
\ee
where the first error is statistical and the second is systematic. 

Effects of the anomalous top quark couplings 
have been widely studied in direct and indirect ways
\cite{larios,lee22,leesong,lee2,lee,jplee,boos}.
Without specifying the underlying model, 
we use an effective lagrangian in this work 
by introducing two complex parameters
such that
\be
{\cal L} = -\frac{g}{\sqrt{2}} 
           \sum_{q=d,s,b}
          V_{tq}^{\rm eff}~ \bar{t} \gamma^\mu
                                 (P_L + \xi_q P_R) q W^+_\mu
  + H.c.,
\ee
where $\xi_q$ are complex parameters measuring effects 
of the anomalous right-handed couplings
while $V_{tq}^{\rm eff}$ measures the SM-like left-handed couplings.
The $\bd$ mixing involves the $tdW$ and $tbW$ couplings
while the $\bs$ mixing involves the $tsW$ and $tbW$ couplings.
On the other hand, the radiative $B \to X_s \gamma$ decay also provides
strict constraints on the $tbW$ and $tsW$ couplings.
If we consider all possible anomalous top quark couplings,
there are too many parameters, 
$3(d,s,b) \times 2 (L-R) \times 2({\rm complex}) = 12$,
and it is hard to get meaningful informations.
Thus we concentrate on the couplings for only one flavour
by keeping the other couplings to be zero.
In the Ref. \cite{lee22}, we have probed the $tsW$ couplings
through $\bs$ mixing and $B \to X_s \gamma$ decay.
We probe the anomalous $tbW$ couplings in this work,
and the $\bd$ mixing should be incorporated
since $tbW$ couplings are common to the $\bd$ and $\bs$ mixings.
Actually the effects of the anomalous right-handed coupling $\xi_b$ 
in $B \to X_s \gamma$ decay are enhanced by $m_t/m_b$ 
due to the structure of the penguin diagram 
in the presence of the right-handed couplings,
but no such enhancements exist for the box diagram. 
Consequently the $\Delta M_q$ constrain only
the anomalous left-handed coupling $V_{tb}^{\rm eff}$, 
while the penguin diagrams constrain both of
$V_{tb}^{\rm eff}$ and $\xi_b$.
Thus the combined analysis of $B-\bar{B}$ mixing
and $B \to X_s \gamma$ decay provides a synergy in
probing the anomalous top couplings.
This paper is organized as follows:
In section II, 
the $B \to X_s \gamma$ constraints on
the anomalous $\bar{t} b W$ couplings is given.
In section III, 
the analysis of the $\bd$ mixing and the $\bs$ mixing 
with anomalous $\bar{t} b W$ couplings is presented. 
Finally we conclude in section IV.

\section{$B \to X_s \gamma$ decays} 

The $\Delta B =1$ effective Hamiltonian for
$b \to s \gamma$ process is given by 
\be
{\cal H}_{eff}^{\Delta B=1} = -\frac{4 G_F}{\sqrt{2}} V_{ts}^* V_{tb} 
           \sum_{i=1}^{8}
              C_i(\mu) O_i(\mu) ,
\ee
where the dimension 6 operators $O_i$ constructed in the SM
are given in the Ref. \cite{buras}.
Matching the effective Hamiltonian and our model given in Eq. (3) 
at $\mu = m_W$ scale,
we obtain the Wilson coefficients $C_i(\mu=m_W)$
\be
C_2(m_W) &=& C_2^{{\rm SM}}(m_W),
\nonumber \\
C_7(m_W) &=& C_7^{{\rm SM}}(m_W) + \xi_b \frac{m_t}{m_b} F_R(x_t),
\nonumber \\
C_8(m_W) &=& C_8^{{\rm SM}}(m_W) + \xi_b \frac{m_t}{m_b} G_R(x_t),
\ee
and otherwise coefficients are zeros, 
where
$C_2(m_W) = -1$,
$C_7(m_W) = F(x_t)$, and
$C_8(m_W) = G(x_t)$
with the well-known Inami-Lim loop functions 
$F(x)$ and $G(x)$ found in
\cite{buras,inami}
and the new loop functions 
\be
F_R(x) &=& \frac{-20+31x-5x^2}{12(x-1)^2}
                 + \frac{x (2-3x)}{2(x-1)^3} \ln x,
\nonumber \\
G_R(x) &=& -\frac{4+x+x^2}{4(x-1)^2}
                 + \frac{3x}{2(x-1)^3} \ln x,
\ee
agree with those in Ref. \cite{cho}.
We note that the anomalous right-handed coupling $\xi_b$ 
involves an enhancement factor $m_t/m_b$.

We obtain the branching ratio of $B \to X_s \gamma$ process
at next-leading-order (NLO) 
in terms of $\xi_b$ as
\be
{\rm Br}(B \to X_s \gamma) &=& {\rm Br}^{\rm SM}(B \to X_s \gamma)
     \left( \frac{|{V_{ts}}^* V_{tb}^{\rm eff}|}
                 {0.0404} \right)^2
     \left[ 1 + Re (\xi_b) \frac{m_t}{m_b} \left(
       0.68 \frac{F_R(x_t)}{F(x_t)} + 0.07 \frac{G_R(x_t)}{G(x_t)} \right)
     \right.
\nonumber \\
     &&~~~~~~~~~~~~   
     \left.
         + |\xi_b|^2 \frac{m_t^2}{m_b^2} 
         \left( 0.112 \frac{F^2_R(x_t)}{F^2(x_t)} 
              + 0.002 \frac{G^2_R(x_t)}{G^2(x_t)} 
              + 0.025 \frac{F_R(x_t) G_R(x_t)}{F(x_t) G(x_t)} \right)
     \right],
\ee
of which numerical coefficients depends on the kinematic cut of
the photon energy spectrum. 
We take the cut $E_\gamma > 1.6$ GeV and the numerical values
are obtained in the Ref. \cite{kagan}.
The SM branching ratio is predicted to be
$ {\rm Br}(B \to X_s \gamma) = (3.15 \pm 0.23) \times 10^{-4} $
with the same photon energy cut at next-to-next-to-leading order (NNLO) 
\cite{bsgammaSM}.
The current world average value
of the measured branching ratio is given by
\cite{HFAG}
\be
{\rm Br}(B \to X_s \gamma) = (3.55 \pm 0.24 ^{+0.09}_{-0.10} \pm 0.03) 
                             \times 10^{-4},
\ee
with the same $E_\gamma$ cut.

\section{$B-\bar{B}$ mixing}

\begin{center}
\begin{figure}[t]
\hbox to\textwidth{\hss\epsfig{file=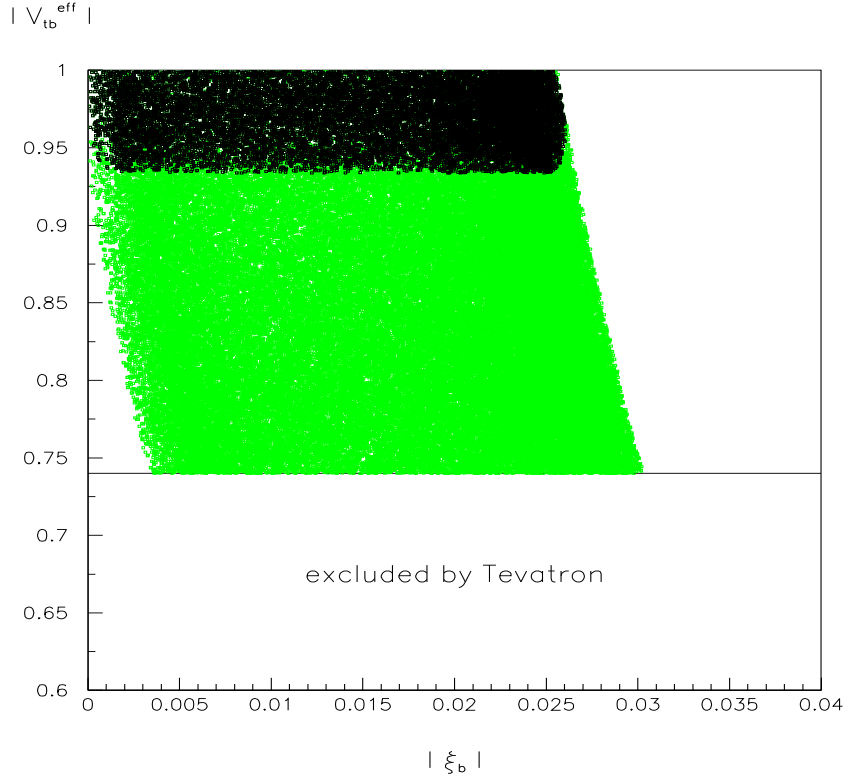,width=13cm,height=11cm}\hss}
 \vskip -1.5cm
\vspace{1cm}
\caption{
Allowed parameter sets $(|\xi_b|, |V_{tb}^{\rm eff}|)$
constrained by $B \to X_s \gamma$ and $\Delta M_s$ (green) and 
by $B \to X_s \gamma$ and $\Delta M_d$ (black).
}
\end{figure}
\end{center}

\begin{center}
\begin{figure}[t]
\hbox to\textwidth{\hss\epsfig{file=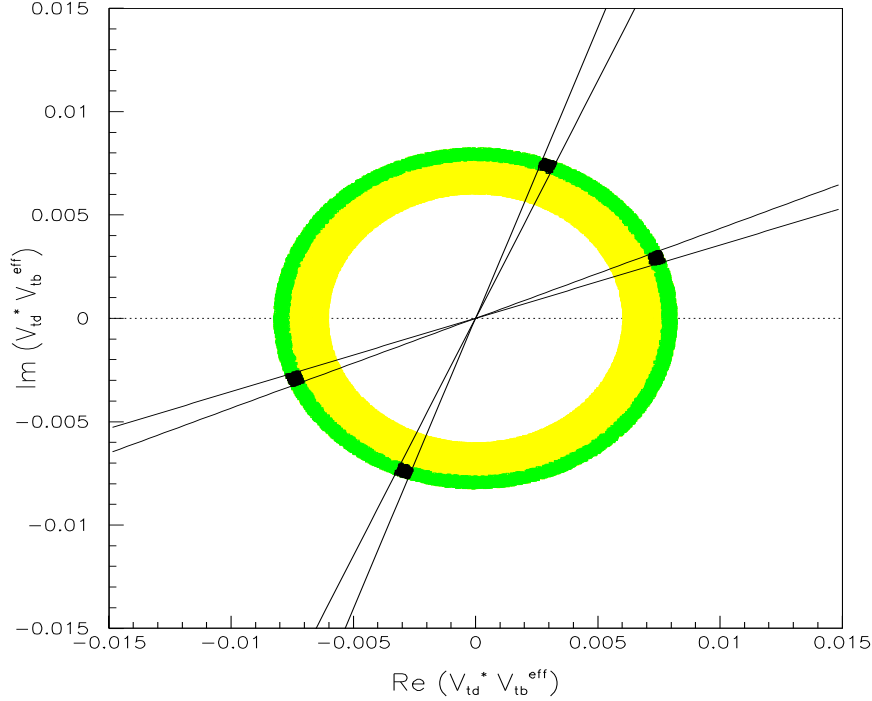,width=13cm,height=11cm}\hss}
 \vskip -1.5cm
\vspace{1cm}
\caption{
The bounds on the complex $V_{td}^* V_{tb}^{\rm eff}$ plane.
The yellow (light gray) region is
allowed by $B \to X_s \gamma$ and $\Delta M_s$
and the green (dark gray) region by $B \to X_s \gamma$ and $\Delta M_d$.
The $\sin 2 \beta$ measurements constrain the phase  
of the $V_{td}^* V_{tb}^{\rm eff}$ with the two-fold ambiguity 
and the allowed regions are denoted by black regions.
}
\end{figure}
\end{center}

A neutral $B_q^0$ meson can oscillate into its antiparticle $\bar{B}_q^0$
via flavour-changing processes of $\bq$ mixing.
The oscillation is described by a Schr\"{o}dinger equation,
\be
i \frac{d}{dt} \left( 
\begin{array}{c} 
B_q(t) \\
\bar{B}_q(t) \\
\end{array}
\right)
= \left( M - \frac{i}{2} \Gamma \right)
\left(
\begin{array}{c} 
B_q(t) \\
\bar{B}_q(t) \\
\end{array}
\right),
\ee
where $M$ is the mass matrix and $\Gamma$ the decay matrix.
The $\Delta B = 2$ transition amplitudes given by
\be
\langle B_q^0 | {\cal H}_{\rm eff}^{\Delta B = 2} | \bar{B}_q^0 \rangle 
         = M_{12}^q,
\ee
is related to the mass difference between the heavy and the light
mass eigenstates,
\be 
\Delta M_q \equiv M_H^{B_q} - M_L^{B_q} = 2 | M_{12}^q |,
\ee
where $M_H^{B_q}$ and $M_L^{B_q}$ 
are the mass eigenvalues for the heavy and the light eigenstates
respectively.
Correspondingly the total decay width difference is defined by
\be 
\Delta \Gamma_q \equiv \Gamma_L^q - \Gamma_H^q.
%                = 2 |\Gamma_{12}^q| \cos \phi_q,
\ee
%for the physical eigenstates $B_L$ and $B_H$
%where the CP phase is given by 
%$\phi_q = {\rm arg}\left( - M_{12}^q/\Gamma_{12}^q \right)$.
The SM predicts the small $\Delta \Gamma_d/\Gamma_d < 1 \%$
and the relatively large $\Delta \Gamma_s/\Gamma_s \sim 10 \%$.
Since the decay matrix elements $\Gamma_{12}^q$ is derived from
the SM decays $b \to c \bar{c} q$ at tree level,
it is hardly affected by the new physics.
We ignore the new effects of the anomalous top couplings 
on $\Delta \Gamma_q$ and just consider the mass differences
in this analysis. 

The box diagrams are calculated to obtain
the transition amplitudes $M_{12}^q$. 
Inclusion of the odd number of right-handed couplings in the box diagram
vanishes due to vanishing the loop integrals of the odd number of momentum.
Thus the leading contribution of the anomalous right-handed top couplings
to the $\bs$ mixing is quadratic order of $\xi_b$.
We write the transition amplitude by
\be
M_{12}^q &=&  \frac{G_F^2 m_W^2}{12 \pi} m_{B_q} 
                 \eta_{q} \hat{B}_{B_q} f_{B_q}^2 S_0(x_t)
     \left(V_{tq}^* V_{tb}^{\rm eff} \right)^2
%                       (\bar{b} \gamma^\mu P_L s) (\bar{b} \gamma_\mu P_L s)
      \left( 1 + \frac{S_3(x_t)}{S_0(x_t)} \frac{{\xi_b^\ast}^2}{4} 
 \frac{\langle B_q^0 |(\bar{b} P_L q) (\bar{b} P_L q)|\bar{B}_q^0 \rangle}
         {\langle B_q^0 |(\bar{b} \gamma^\mu P_L q) 
                         (\bar{b} \gamma_\mu P_L q)|\bar{B}_q^0 \rangle}
\right),
%\nonumber \\
%&\equiv& M_{12}^{\rm SM,q} \cdot \Delta_q,
\ee
where $\eta_q$ are the perturbative QCD corrections 
to the $\bq$ mixings \cite{QCD}.
The Inami-Lim loop functions are given by
\be
S_0(x) &=& \frac{4x-11 x^2 + x^3}{4 (1-x)^2} - \frac{3 x^3}{2 (1-x)^3} \log x,
\nonumber \\
S_3(x) &=& 4 x^2 \left( \frac{2}{(1-x)^2} + \frac{1+x}{(1-x)^3} \log x \right).
\ee
Vacuum insertions to the hadronic matrix elements lead to
\be
\frac{\langle B_q^0 |(\bar{b} P_L q) (\bar{b} P_L q)|\bar{B}_q^0 \rangle}
         {\langle B_q^0 |(\bar{b} \gamma^\mu P_L q) 
                         (\bar{b} \gamma_\mu P_L q)|\bar{B}_q^0 \rangle}
        = \frac{5}{8} \left( \frac{m_{B_q}}{m_b+m_q} \right)^2 ,
\ee
and 
\be
\langle B_q^0 | (\bar{b} \gamma^\mu P_L q) (\bar{b} \gamma_\mu P_L q)
 | \bar{B}_q^0 \rangle = \frac{8}{3} m_{B_q}^2 \hat{B}_{B_q} f_{B_q}^2 ,
\ee
where $\hat{B}_{B_q}$ are the Bag parameters 
and $f_{B_s}^2$ the decay constants.
The SM predictions of the $B-\bar{B}$ mixings are 
given by $\Delta M_s = 19.3 \pm 6.74$ ps$^{-1}$ 
and $\Delta M_d = 0.53 \pm 0.02$ ps$^{-1}$ 
\cite{nierste}.

We show the allowed parameter sets $(|\xi_b|, |V_{tb}^{\rm eff}|)$ 
at 95\% C. L. in Fig. 1. 
The black region is allowed by 
${\rm Br}(B \to X_s \gamma)$ and $\Delta M_d$
while the green (gray) region allowed by
${\rm Br}(B \to X_s \gamma)$ and $\Delta M_s$.
We have the conservative bounds
$|V_{tb}| > 0.93$ and $|\xi_b|<0.027$ in Fig. 1.
Since $\xi_b$ and $V_{tb}^{\rm eff}$ are complex parameters, 
the new physics effects 
arise in both magnitude and phase of $M_{12}^q$ in general.
Effects of the phase and CP violation in $M_{12}^s$ 
have been measured \cite{cpexp}, although not very accurately, 
and discussed in several literatures 
\cite{nierste,cp}.
The CP phase of the $\bd$ mixing is measured 
through the $B \to J/\psi K_s$
and has been tested in many $B$ decay processes
\cite{BBcp}.
The recent world average value of the weak phase defined by
$\sin 2 \beta = -(V_{td} V_{tb}^*)/(V_{td}^* V_{tb})$
is given by \cite{HFAG}
\be
\sin 2 \beta = 0.680 \pm 0.025
\ee
through the time-dependent CP asymmetries into all charmonium states.
Figure 2 shows the allowed values of
$V_{td}^* V_{tb}^{\rm eff}$ on the complex plane.
The yellow (light gray) region denotes the allowed region 
by $B \to X_s \gamma$ and $\Delta M_s$,
and the green (dark gray) region by $B \to X_s \gamma$ and $\Delta M_d$.
The allowed region by the $\sin 2 \beta$ measurements has
the two-fold ambiguity on the complex $V_{td}^* V_{tb}^{\rm eff}$ plane.
The black region denotes the allowed regions 
additionally by the world average values of $\sin 2 \beta$ measurements.

%Current experimental data on the phase of $M_{12}^s$ 
%show some deviations from the SM prediction. 
%Figure 2 shows the scanned parameters in the complex $\Delta_s$ plane.
%The box is the combined experimental data given in Ref. \cite{nierste}.
%Note that the phase of $\Delta_s = M_{12}^s/M_{12}^{\rm SM,s}$
%is dominated by the phase of $V_{ts}^{\rm eff}$ in our model.

\section{Concluding Remarks}

The neutral $B_q^0$ meson systems are of great use 
for search for the new physics effects in top quark couplings.
We consider the anomalous $tbW$ couplings
parametrized by $V_{tb}^{\rm eff}$ and $\xi_b$.
Combined analysis of $\bs$ mixing, $\bd$ mixing 
and $B \to X_s \gamma$ penguin decay
provides strong constraints on the parameters of
$V_{tb}^{\rm eff}$ and $\xi_b$.
We find that the bounds from $\bd$ mixing 
is better than that from $\bs$ mixing.
It is because the SM prediction of $\Delta M_d$
is more precise than that of $\Delta M_s$.

\acknowledgments
This work was supported by
the Korea Research Foundation Grant funded by the Korean Government
(MOEHRD, Basic Research Promotion Fund KRF-2007-C00145)
and the BK21 program of Ministry of Education (K.Y.L.).

%%%%%%%%%%%%%%%%%% References
%%%%%%%%%%%%%%%%%%%%%%%%%%%%%%%%%%%%%%%%%%%%%%%%%%%%%%%%%%%%%%%%%%%%%%%
\def\PRD #1 #2 #3 {Phys. Rev. D {\bf#1},\ #2 (#3)}
\def\PRL #1 #2 #3 {Phys. Rev. Lett. {\bf#1},\ #2 (#3)}
\def\PLB #1 #2 #3 {Phys. Lett. B {\bf#1},\ #2 (#3)}
\def\NPB #1 #2 #3 {Nucl. Phys. {\bf B#1},\ #2 (#3)}
\def\ZPC #1 #2 #3 {Z. Phys. C {\bf#1},\ #2 (#3)}
\def\EPJ #1 #2 #3 {Euro. Phys. J. C {\bf#1},\ #2 (#3)}
\def\JHEP #1 #2 #3 {JHEP {\bf#1},\ #2 (#3)}
\def\IJMP #1 #2 #3 {Int. J. Mod. Phys. A {\bf#1},\ #2 (#3)}
\def\MPL #1 #2 #3 {Mod. Phys. Lett. A {\bf#1},\ #2 (#3)}
\def\JPG #1 #2 #3 {J. Phys. G {\bf#1},\ #2 (#3)}
\def\PTP #1 #2 #3 {Prog. Theor. Phys. {\bf#1},\ #2 (#3)}
\def\PR #1 #2 #3 {Phys. Rep. {\bf#1},\ #2 (#3)}
\def\RMP #1 #2 #3 {Rev. Mod. Phys. {\bf#1},\ #2 (#3)}
\def\PRold #1 #2 #3 {Phys. Rev. {\bf#1},\ #2 (#3)}
\def\IBID #1 #2 #3 {{\it ibid.} {\bf#1},\ #2 (#3)}
%%%%%%%%%%%%%%%%%%%%%%%%%%%%%%%%%%%%%%%%%%%%%%%%%%%%%%%%%%%%%%%%%%%%%%%

\end{document}